\documentclass[conference]{IEEEtran}
\IEEEoverridecommandlockouts
\usepackage{cite}
\usepackage{amsmath,amssymb,amsfonts}
\usepackage{graphicx}
\usepackage{textcomp}
\usepackage{xcolor}
\def\BibTeX{{\rm B\kern-.05em{\sc i\kern-.025em b}\kern-.08em
    T\kern-.1667em\lower.7ex\hbox{E}\kern-.125emX}}

\usepackage{bbding}

\usepackage{enumitem}
\usepackage{epstopdf}
\usepackage{subfigure}
\usepackage{multirow}
\usepackage[ruled,linesnumbered]{algorithm2e}

\usepackage{balance}

\usepackage[hyphens]{url}
\usepackage[hidelinks]{hyperref}
\usepackage{float}
\newcommand{\myparagraph}[1]{\vspace{0.3\baselineskip}\noindent{\textbf{#1.}}~}
\newcommand\systemname{\textsc{{Ubis}}}
\newcommand\repourl{https://github.com/whu-totemdb/UBIS.git}

\DeclareRobustCommand*{\IEEEauthorrefmark}[1]{%
    \raisebox{0pt}[0pt][0pt]{\textsuperscript{\footnotesize\ensuremath{#1}}}}

\begin{document}

\title{Updatable Balanced Index for Stable Streaming Similarity Search over Large-Scale Fresh Vectors
}


\author{
\IEEEauthorblockN{
Yuhui Lai\IEEEauthorrefmark{1},
Shixun Huang\IEEEauthorrefmark{2},
Sheng Wang\IEEEauthorrefmark{1\textsuperscript{*}}\thanks{\textsuperscript{*} Sheng Wang is the corresponding author.}
}

\IEEEauthorblockA{\IEEEauthorrefmark{1}School of Computer Science, Wuhan University, Wuhan, China}
\IEEEauthorblockA{\IEEEauthorrefmark{2}School of Computing and Information Technology, The University of Wollongong, Wollongong, Australia}
\IEEEauthorblockA{huifei@whu.edu.cn, shixunh@uow.edu.au, swangcs@whu.edu.cn}
}

\maketitle

\begin{abstract}
As artificial intelligence gains more and more popularity, vectors are one of the most widely used data structures for services such as information retrieval and recommendation. Approximate Nearest Neighbor Search (ANNS), which generally relies on indices optimized for fast search to organize large datasets, has played a core role in these popular services. As the frequency of data shift grows, it is crucial for indices to accommodate new data and support real-time updates. Existing researches adopting two different approaches hold the following drawbacks: 1) approaches using additional buffers to temporarily store new data are resource-intensive and inefficient due to the global rebuilding processes; 2) approaches upgrading the internal index structure suffer from performance degradation because of update congestion and imbalanced distribution in streaming workloads. In this paper, we propose {\systemname}, an \underline{U}pdatable \underline{B}alanced \underline{I}ndex for stable streaming similarity \underline{S}earch, to resolve conflicts by scheduling concurrent updates and maintain good index quality by reducing imbalanced update cases, when the update frequency grows. Experimental results in the real-world datasets demonstrate that {\systemname} achieves up to 77\% higher search accuracy and 45\% higher update throughput on average compared to the state-of-the-art indices in streaming workloads. 
\end{abstract}


\section{Introduction}

With the rapid growth of unstructured data \cite{king80percent,li2021adsgnn} such as images, videos, and documents, the challenges of data management have become increasingly significant. In recent years, the widespread adoption of machine learning models \cite{le2014distributed,barkan2016item2vec} has allowed the transformation of unstructured data into feature vectors. These generated high-dimensional vectors are utilized to perform Approximate Nearest Neighbor Search (ANNS). It aims to search the closest vectors in the datasets and has wide applications, such as content retrieval \cite{li2021embedding} and recommendation \cite{wei2020analyticdb,covington2016deep}.

\begin{figure}
\setlength{\abovecaptionskip}{-0.3 cm}
\setlength{\belowcaptionskip}{0 cm}
    \centering
    \includegraphics[width=\linewidth]{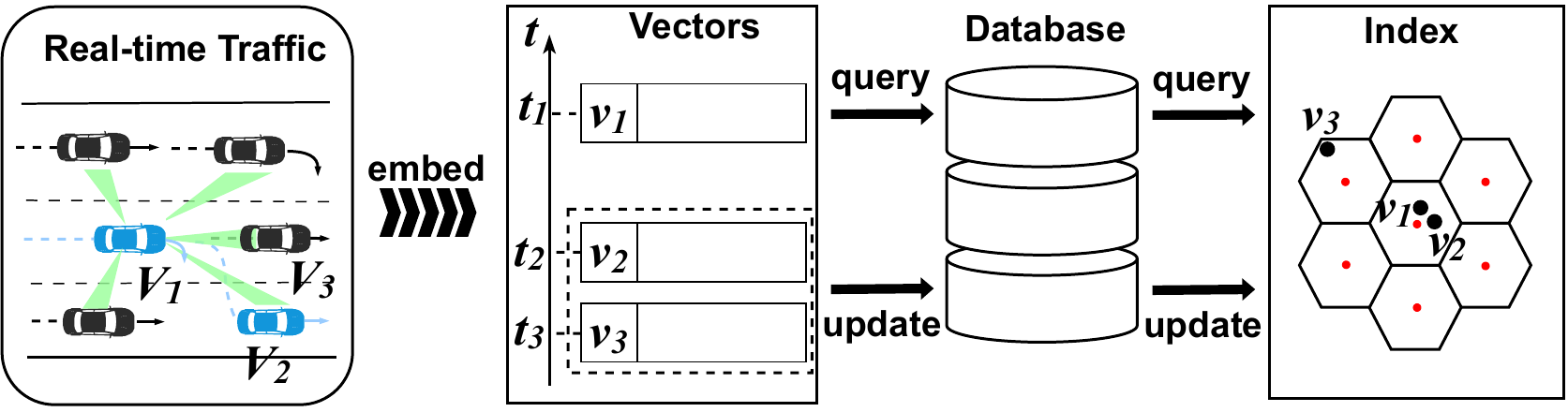}
 \caption{A real-world scenario: trajectories of vehicles $V_i$ are embedded as feature vectors $v_i$ with timestamp $t_i$, where similarity search and update are performed concurrently. A real-time updatable index requires receiving fresh vectors in time such that the query vector $v_1$ can access the latest $v_2$.}

    \label{fig:real-world scenarios}
\end{figure}

\myparagraph{Dynamic Search Demands} Nowadays ANNS has become a cornerstone service for the state-of-the-art vector search systems \cite{wang2021milvus,li2018design,douze2024faiss,jayaram2019diskann}. They rely on an index to organize these vectors in the datasets to accelerate the search process. However, more and more applications require updatable indices to provide real-time ANNS. Users generally prefer to receive the latest information in the endless new data flows, and stale data may disturb the quality of search results. For instance, e-commerce platforms \cite{li2021embedding} require upgrading the data pools to help customers quickly search their desired new products, users of social media \cite{x2022recommendation} are more willing to receive the latest blogs and posts aligned with their interests, and Large Language Models (LLMs) \cite{yu2025topology,pound2025micronn} also need instant updates on the index to retrieve relevant fragments from chat histories as new prompts. Autonomous driving \cite{ji2024simplifying, Argoverse2} is also a typical application. Trajectories can be obtained through images from surveillance cameras \cite{zuo2024serf,zuo2023arkgraph} and detection devices \cite{van2018autonomous}. Autonomous vehicles need to quickly search for the most similar trajectories to predict future conditions and plan a safe route \cite{wangkdd2022}. Based on experience, the most similar historical trajectory data is highly likely to be a safer path for the autonomous vehicle. As shown in Figure \ref{fig:real-world scenarios}, the autonomous vehicle $V_1$ attempts to plan a safe route. It performs a similarity search based on its own real-time trajectory vector $v_1$ to quickly search for a historical route in the database, while the database is updated by the real-time trajectories of other vehicles, such as $v_2$ and $v_3$. If the ANNS system fails to update the index in time, the most valuable vector $v_2$ cannot contribute to the driving decision of $V_1$.

\myparagraph{Existing Indices and Update Approaches}
Existing indices can be primarily divided into two major categories: \textit{cluster-based} \cite{chen2021spann,douze2024faiss} and \textit{graph-based} \cite{malkov2018efficient,malkov2012scalable}. Cluster-based indices store the vectors within different clusters, where vectors that share more features are in the same group with higher probability. Graph-based indices explicitly connect those similar vectors through the neighborhood relationships (details in Section \ref{sec:index structure}). As more and more applications demand fast and dynamic vector search services, the index update issue has been attracting attention in recent years. Most existing update solutions \cite{li2018design,singh2021freshdiskann,ono2023relative} implement index update through the index \textit{rebuild} process, which accumulates new vectors in the buffer and rebuilds the index from scratch. 
They construct additional new indices to adapt to changes in the datasets, called \textit{out-of-place update}. SPFresh \cite{xu2023spfresh} is the state-of-the-art method in bridging the gap of updating the internal index structure, called \textit{in-place update}. To efficiently evolve the cluster-based index with new data, SPFresh employs a new policy that clusters are adaptively adjusted with low costs (details in Section \ref{sec:index update issue}).

\myparagraph{{Drawbacks of Existing In-place Update Approaches}} In this paper, we focus on \textit{streaming workloads} \cite{aguerrebere2024locally,singh2021freshdiskann,mohoney2024incremental,xu2025place}, where update requests are processed and search tasks are required to be completed in time. SPFresh, the representative of in-place update approaches, has the following two drawbacks in streaming workloads:

\begin{itemize}
[leftmargin=1em]
    \item[1)] It fails to update the index in a stable way, since multiple update jobs are concurrently executed, which has the risk of data contentions and conflicts. Fewer fresh vectors are added to the index when update jobs are interrupted.
    \item[2)] It fails to check some update cases where some clusters of small sizes are generated. A narrow search with fewer accessed vectors occurs when the distribution is imbalanced.
\end{itemize}

\vspace{-0.2cm}
\myparagraph{Our Solution} To the best of our knowledge, it is still non-trivial for updating index under streaming workloads with high update frequency. To mitigate existing drawbacks above, in this paper we propose {\systemname}, an \underline{U}pdatable \underline{B}alanced \underline{I}ndex for stable streaming similarity \underline{S}earch under streaming workloads, to address the following two main challenges: 1) To keep updating the index efficiently and perform similarity search stably, {\systemname} needs to resolve the potential execution conflicts in the competitive streaming workloads. 2) To eliminate the impact of streaming updates on data distribution, {\systemname} requires maintaining good index quality and preventing imbalanced update cases as the data shifts and the centroids vary. {\systemname} achieves the goals by observing that different update operations usually hold different operated objects and expected results. For example, only the newly-generated clusters need to be paid close attention to, because the old ones are eventually discarded, and insertions only focus on the expansion of the original clusters. Therefore, these jobs can be executed more efficiently because they cannot block each other.

\par In summary, we make the following contributions:

\begin{itemize}
[leftmargin=1em]

    \item We design a novel structure to record the latest update states, and propose a new mechanism by scheduling different updates, in a fine-grained manner, to resolve data contentions and stably update the internal index structure (Section \ref{sec:fine-grained concurrency control}).
    \item We enhance the ability to maintain a more balanced distribution in streaming workloads by relaxing constraints and identifying imbalanced update cases (Section \ref{sec:balanced online update}).
    
    \item We are the first to test update performance on both data-driven datasets with real-world timestamps and synthetic modeling datasets with simulated orders (Section \ref{sec:experiment setup}).
    
    \item Experiments on real-world streaming workloads demonstrate that {\systemname} achieves up to 77\% higher search accuracy with 45\% higher update efficiency compared to the state-of-the-art ANNS indices (Section \ref{sec:streaming update experiment}).

\end{itemize}

\section{Related Work}

\begin{table}[t]
\centering
\setlength{\abovecaptionskip}{-0.01 cm}
\caption{Update comparison to cluster-based ANNS systems.}
\label{tab:Comparison with cluster-based ANNS systems}

\resizebox{0.9\linewidth}{!}{
\begin{tabular}{cccc}
\hline
        & Increment & Stream & Balance \\
\hline
SPANN\cite{chen2021spann}   &         \XSolidBrush           &        \XSolidBrush          &        \XSolidBrush         \\
SPFresh\cite{xu2023spfresh} &          \CheckmarkBold          &             \XSolidBrush     &           \XSolidBrush      \\
{\systemname}    &   \CheckmarkBold                 &       \CheckmarkBold           &      \CheckmarkBold           \\ \hline
\end{tabular}
}

\end{table}

In this section, we briefly introduce the index structures used in existing ANNS systems, followed by studies on index update. Afterwards, differences between {\systemname} and other state-of-the-art ANNS methods are introduced.

\subsection{Existing Index Structures}
\label{sec:index structure}

\myparagraph{Cluster-based Index} Cluster-based indices partition the dataset into clusters (also called postings), and organize vectors hierarchically or flatly to accelerate search processes. Hashing algorithms \cite{he2010scalable,sun2014srs,datar2004lsh,gionis1999similarity} and clustering algorithms \cite{lloyd1982least,muja2014scalable,wang2020efficiency,f3km,ji2025vldb} are widely used to divide different partitions. This kind of index, such as SPANN \cite{chen2021spann} (details in Section \ref{sec:index structure}), typically relies on a two-phase search process: coarse-grained filtering to get close clusters and fine-grained searching within the selected clusters.

\myparagraph{Graph-based Index} Graph-based indices construct a proximity graph to organize all vectors in the datasets. Based on the idea that a neighbor's neighbor is likely to be a neighbor, graph-based ANNS indices \cite{malkov2012scalable,malkov2018efficient,jayaram2019diskann,ren2020hm} utilize edges to represent the neighbor relationship between two vectors with high similarity. The search process is generally performed by starting from an entry vector and navigating to neighbors under the greedy traversal policy.

\subsection{Existing Index Update Methods}
\label{sec:index update issue}

Generally, indices are rebuilt to receive new data because direct modifications bring high costs and result in poor search performance, which is inefficient. Cluster-based index is more friendly to receive fresh data than the graph-based index where complex neighborhoods need to be maintained.

\myparagraph{Out-of-place Update} Methods in this category \cite{li2018design,singh2021freshdiskann,jayaram2019diskann} leverage external buffers to store new vectors. Then these vectors are merged into the datasets, the old index is discarded, and a new index is rebuilt, which is resource-intensive.

\myparagraph{In-place Update} Methods in this category directly modify the index structure to implement incremental update. SPFresh \cite{xu2023spfresh} is the state-of-the-art incremental in-place update solution for SPANN. It splits the oversized clusters and merges the small clusters to shift centroids in local regions (details in Section \ref{sec:update policy}). Ada-IVF \cite{mohoney2024incremental} employs a lazy update strategy through access frequency. Hot clusters are more likely to become candidates for updating. IP-DiskANN \cite{xu2025place} supports in-place deletion for graph-based index to reduce update costs.

\myparagraph{Summary} These researches are not suitable for streaming workloads, because rebuilding the index introduces time costs and existing incremental solutions fail to efficiently update and maintain a more balanced distribution. In this paper, we focus on cluster-based index and compare {\systemname} with the representative cluster-based ANNS systems, as shown in Table \ref{tab:Comparison with cluster-based ANNS systems}. SPFresh solves the incremental update problem in SPANN, while {\systemname} enhances the ability to stream updates and balanced updates that are not resolved in these indices.

\section{Preliminary}

In this section, we formulate the streaming update problem for cluster-based index. We briefly present the SPANN index, which is the basis of our method, and SPFresh's update policy with mathematical description before introducing {\systemname}. 

\begin{figure}
    \centering
    \setlength{\abovecaptionskip}{-0.1 cm}
\setlength{\belowcaptionskip}{0.2cm}
\includegraphics[height=0.2\textwidth]{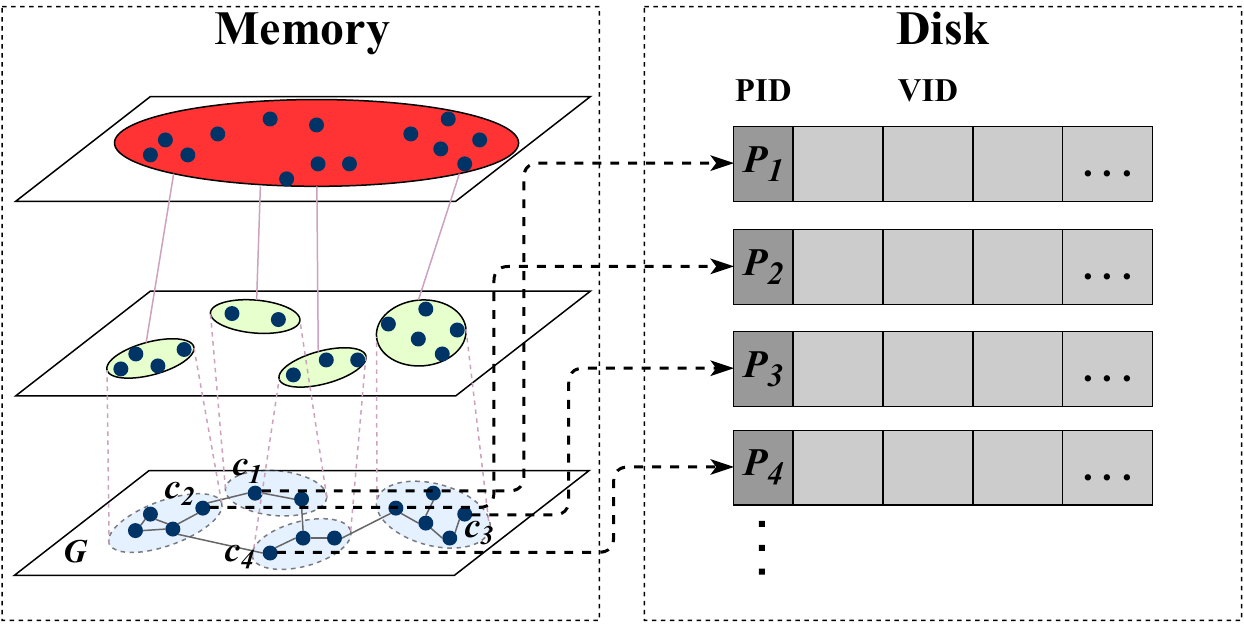}
    \caption{SPANN index structure.}  
    \label{fig:spann index structure}
\end{figure}

\subsection{Notations}

The cluster-based index consists of lots of postings, where each posting stores the vectors that are closest to the centroid. 

\myparagraph{Basic Concepts}We let $D$ denote the dataset that is waiting to be indexed, where $|D|=n$. A posting $P_i$ with its own identifier $i$ consists of a unique centroid $c_i$ and a collection of $|P_i|$ vectors, where $\forall p \in P_i, p \in D$. Therefore, we let $I=\{P_1, P_2, ..., P_{|I|}\}$ denote the cluster-based index with $|I|$ postings where $n = \sum_{i=1}^{|I|}|P_i|$. We denote the distance between two vectors $p$ and $q$ by the function $d(p,q)$. Generally, $d(p,q)=\lVert p-q \rVert$, which denotes the Euclidean distance, is one of the widely used functions to evaluate distance.

\myparagraph{$k$-NN Search} Given a query vector $q$ and the number of the nearest neighbors $k$, $k$-NN search, denoted by $R_k(q)$, aims to return a collection $R$ containing the top $k$ nearest vectors in the dataset $D$ of the query $q$ using index $I$ based on distance function $d$, which is formulated as followed:
\begin{equation}
    R = \mathop{\arg\min}_{|S|=k \bigwedge S \subseteq D} \sum_{i=1}^{k} d(p,q), p \in S.
\end{equation}

\noindent The search accuracy is reflected by the ratio $\frac{|R \cap T|}{|T|}$, named recall rate, where $T$ is the ground truth set.

\myparagraph{Problem Formulation (Cluster-based Index Streaming Update Problem)}The initial cluster-based index $I_0$ is built based on the initial dataset $D_0$ with $|D_0|$ vectors. Given $m$ new datasets to be appended $D_1, D_2,...,D_m$, we aim to update the existing index $I_{j-1}, j \in [1,m]$, using dataset $D_j$ in the $j$-th batch, in an efficient and effective way such that $k$-NN search can be performed on the evolved index $I_j$ to maintain high accuracy with less latency.

\subsection{In-place Update for SPANN index} 
\subsubsection{Index Structure}
\label{sec:spann index structure}SPANN \cite{chen2021spann} is the state-of-the-art cluster-based index, which utilizes a Space Partition Tree And Graph (SPTAG) to search for the nearest vectors in large-scale datasets. As Figure \ref{fig:spann index structure} shows, SPANN index $I=\{P_1, P_2, ..., P_{|I|}\}$ organizes the postings $P_i \in I$ in a hierarchical structure, named Balanced $k$-Means Tree (BKT). In addition, it maintains a relative neighborhood graph $G=(C,E)$ in the bottom layer, where $C=\{c_1,c_2,...,c_{|I|}\}$, $c_i$ is the centroid of $P_i$ and $e=(c_i, c_j) \in E$ represents the neighborhood relationship of these two centroids, if they share a high similarity. Each vector $p$ is in the group whose centroid $c_i$ is closest to itself, which means that the vector $p$ satisfies $d(p,c_i) \leq d(p, c_j), \forall c_j \in C$.

\subsubsection{Update Policy}\label{sec:update policy}SPFresh \cite{xu2023spfresh} employs a LIRE protocol on SPANN to support online update. It proposes three new operations to handle the data shift and maintain index quality:

\begin{itemize}[leftmargin=1em]
    \item \textbf{Split.} Given the maximum of the posting size $l_{max}$, the split operation is triggered when new vectors are inserted into the index $I$ and the target posting $P_i \in I$ satisfies the condition where $|P_i| > l_{max}$. The original posting $P_i$ is split into two smaller sub-postings $P_{i1}$ and $P_{i2}$, where $|P_i|=|P_{i1}|+|P_{i2}|$. Finally, a reassignment operation is triggered.
    \item \textbf{Merge.} Given the minimum and the maximum of the posting size $l_{min}$ and $l_{max}$, merge operation is triggered, when performing vector nearest search tasks in the index $I$ and the target posting $P_{i1} \in I$ satisfies the condition where $|P_{i1}| < l_{min}$. The nearest posting $P_{i2}$ that satisfies $|P_{i1}|+|P_{i2}|< l_{max}$ is the candidate to merge with $P_{i1}$. The new posting $P_i$ is the combination of these two postings $P_{i1}$ and $P_{i2}$, where $|P_i|=|P_{i1}|+|P_{i2}|$. Finally, a reassignment operation is triggered.
    \item \textbf{Reassign.} Given a posting $P$, the reassignment operation is triggered by split or merge operations. It traverses each vector $p \in P$ to check whether there is a nearer posting $P'$ that satisfies $d(p,c) > d(p,c')$, where $c$ is the centroid of $P$ and $c'$ is the centroid of $P'$.
\end{itemize}

\section{{\systemname}}
\label{sec:system design}

Original SPANN index can't be updated incrementally due to lower search accuracy. SPFresh proposes a local update strategy to solve it. But SPFresh still faces performance decline in streaming workloads due to contentions and imbalances. Therefore, we propose {\systemname} to overcome these drawbacks.
In this section, we will describe the detailed design of {\systemname} from the following aspects:
\begin{itemize}
[leftmargin=1em]
    
    \item We provide an overview of new modules that help to implement an updatable balanced index in streaming workloads, and describe their functions in general (Section \ref{sec:overview}).
    \item We present the details of the fine-grained concurrency control mechanism and the underlying structure, to achieve effective updates with fewer contentions (Section \ref{sec:fine-grained concurrency control}). 
    \item Based on the discovery of the distribution experiment, we incorporate additional update branches to identify imbalanced cases, which helps to achieve a more uniform distribution and get more accurate results (Section \ref{sec:balanced online update}).
\end{itemize}

\subsection{Overview}\label{sec:overview}

\begin{figure}
    \centering
     \setlength{\abovecaptionskip}{-0.5 cm}
\setlength{\belowcaptionskip}{0.3 cm}
\includegraphics[width=\linewidth]{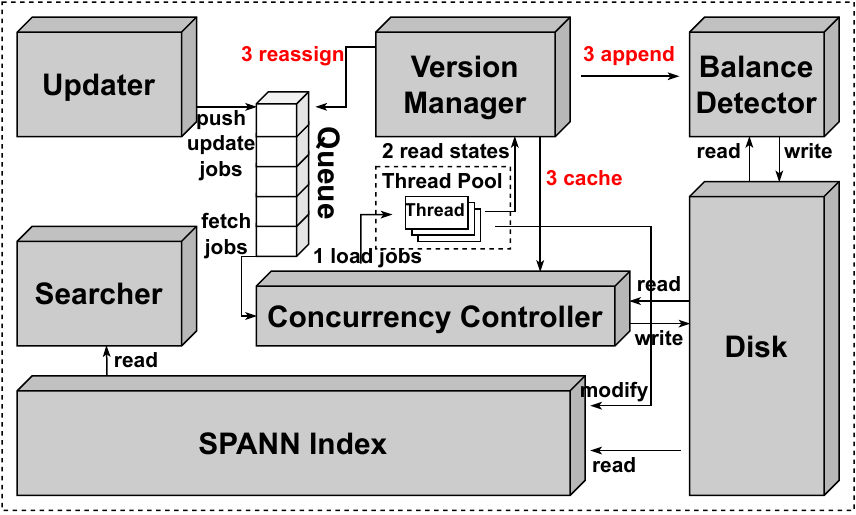}
    \caption{The overall architecture of {\systemname}. The red texts mean there are three branches for the third step, the decision depends on status in \textit{version manager}.}  
    \label{fig:our architecture}
\end{figure}
 
Figure \ref{fig:our architecture} shows the overall architecture of our proposed updatable balanced index {\systemname}. Considering the complexity of internal structure modification and the frequency of update operations, {\systemname} requires to guarantee that updates maintain index quality as much as possible, and these jobs can be executed in parallel with fewer potential barriers. To achieve this, we propose new components, called a \emph{fine-grained version manager}, a \emph{high-concurrency controller}, and a \emph{balance detector}, to implement an updatable balanced index, which helps to handle the complex streaming workloads in real-world applications and scenarios. As for the storage engine, we select RocksDB \cite{dong2021rocksdb} to persist data on disk.

\myparagraph{Fine-Grained Version Manager}The fine-grained version manager is designed to manage data updates at a granular level, ensuring that changes are tracked and applied efficiently. This module allows for more precise control over individual postings. By maintaining detailed version histories, the index can quickly obtain previous states if necessary, reducing the risk of data corruption in the streaming update workloads.

\begin{figure*}[ht]
    \centering
    \setlength{\abovecaptionskip}{-0.2 cm}
\setlength{\belowcaptionskip}{-0.5 cm}
\includegraphics[height=0.18\textwidth]{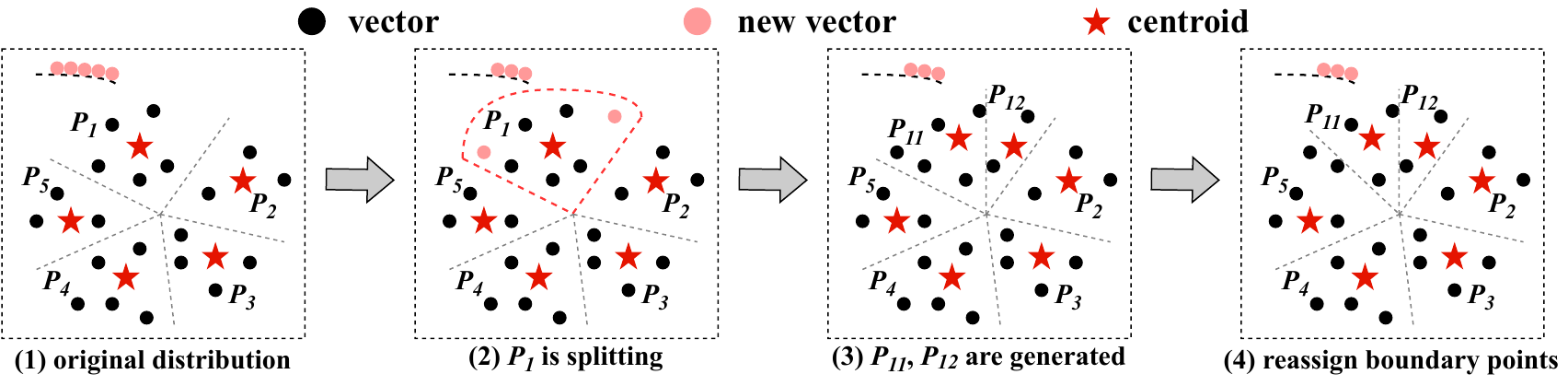}
    \caption{An example of in-place update in SPFresh.}  
    \label{fig:an example of in-place update for cluster-based index}
\end{figure*}

\myparagraph{High-Concurrency Controller}The high-concurrency controller is engineered to handle multiple concurrent update requests. Concurrent updates can lead to potential conflicts. This module employs fine-grained mechanisms based on the data in \textit{fine-grained version manager}, to ensure that updates are processed correctly and data contentions are minimized.

\myparagraph{Balance Detector}The balance detector is responsible for ensuring that the index stays balanced during multiple updates. Generally, frequent updates can lead to uneven data distribution over time, which can degrade performance. This module monitors the distribution of the data and analyzes the imbalanced update cases. Rebalance operations are triggered when successfully detecting imbalanced situations. By proactively addressing imbalances, the balance detector helps to sustain the efficiency and accuracy of the index.

\myparagraph{Workflow}As shown in Figure \ref{fig:our architecture}, update jobs are pushed into the job queue when receiving fresh vectors. The concurrency controller fetches these jobs and loads them with multiple threads. Each job will locate their target postings and read their versions stored in the version manager. Based on the information, fresh vectors are processed in different ways. When an old posting is to be split or merged, the balance detector will check whether new postings are balanced. If new postings are imbalanced, reassignments will be triggered.

\subsection{Fine-grained Concurrency Control}\label{sec:fine-grained concurrency control}
SPFresh suffers from update contention during several former update batches due to its coarse-grained posting-level lock design. As shown in Figure \ref{fig:an example of in-place update for cluster-based index}, the new vectors must be blocked until the posting ends splitting. If a deleted posting is identified, SPFresh will repeat the search process. It will search the neighbors of the deleted posting, select the nearest one and append new vectors to it, which degrades the search accuracy and update efficiency.
\par To resolve the problem, we design a fine-grained concurrency control mechanism to reduce the potential contention and the additional search for near postings. The implementation benefits from the newly-proposed fine-grained version manager and the newly-proposed high-concurrency controller, which helps to control different concurrent jobs in a fine-grained way. The version manager records the states of the involved postings during the update operations, while the high-concurrency controller schedules different update tasks depending on the recorded states.

\subsubsection{Fine-Grained Version Manager} Generally, each update job described in Section \ref{sec:update policy} is executed in a feed-forward pipeline, where the foreground feeds the job into a queue and the background fetches each job in the queue. However, SPFresh fails to manage the versions of the evolving postings generated by these update jobs during this process. Therefore, it needs to leverage the posting-level lock to guarantee the correctness of its proposed write operations, which is the potential bottleneck of performance in competitive streaming workloads. Based on the discovery, we propose a new data structure, called \textit{Posting Recorder}, in the version manager to record and manage the updating states of the postings.

\par We utilize a dense array in memory to store the information. In order to efficiently handle the versions of all postings while not introducing too many storage overheads, the state information of each posting takes 8 bytes by default. Each entry consists of the following parts:

\begin{itemize}[leftmargin=1em]
    \item \textbf{Status.} The status data records the posting state. In the context of the cluster-based index, the status data for a single posting is efficiently represented using only 2 bits, since there are four possible states: the posting is in a normal state, the posting is undergoing splitting, the posting is undergoing merging, or the posting is marked as deleted.
    \item \textbf{Weight.} This region takes 16 bits by default and contains the version information. It marks the visibility of the posting. A search is based on a snapshot, and a posting is visible for the current snapshot, only when the global version is greater than the weight of the posting.
    \item \textbf{New Postings.} This region utilizes the remaining bits to encode the identifiers of the new postings generated during split or merge operations, which means that this region represents the pointers to the new postings.
\end{itemize}

\par To summarize, status stores a posting's state, weights record the versions to check if a search can access the posting based on global versions, and new postings store the identifiers of newly generated postings to quickly locate them. We use the compare-and-swap (CAS) operation to modify the data in \textit{Posting Recorder}, so atomicity can be guaranteed.

\subsubsection{High-Concurrency Controller}This new module facilitates the implementation of an updatable index, allowing for the concurrent execution of different update operations. On the basis of introducing our proposed \textit{Posting Recorder}, the new high-concurrency controller can eliminate the posting-level lock design in SPFresh and reduce the data conflicts. 

\par Based on the recorded status of the corresponding entry in the \textit{Posting Recorder}, our proposed high-concurrency controller can properly schedule concurrent jobs in the job queue. There are three cases for the state of the target posting, which is displayed in Figure \ref{fig:our architecture} for the third step of each update job:

\begin{itemize}[leftmargin=1em]
    \item \textbf{Normal Posting.} In this case, the process of each job is similar to that in SPFresh. The new vectors will be appended to the posting or the versions of the target vectors will be modified, and the target vectors will be marked as deleted. Afterwards, the \textit{Balance Detector} will also check the updated cases to prevent some imbalanced cases.
    \item \textbf{Deleted Posting.} When the target posting is marked as deleted, the controller will first check whether there exist alive sub-postings or alive parent postings. If so, the controller will append the new vectors to the posting or attempt to search the target vectors in the posting. Otherwise, it will trigger a reassign job and push it to the queue. A reassign job will append the available vectors to the posting that is closest to the old deleted posting.
    \item \textbf{Splitting or Merging Posting.} The controller maintains a vector cache that is optimized for high-concurrency context. The new vectors during the time-cost split or merge processes will be stored in the vector cache. These vectors can be accessed by the search tasks like other normal vectors. When the split or merge process is completed, the vectors in the cache will be appended to the nearest new posting.
\end{itemize}

\par To summarize, our proposed fine-grained controller can get rid of the locks at posting granularity in the design of SPFresh and resolve some potential conflicts by leveraging the scheduling strategy.

\subsection{Balanced Online Update}\label{sec:balanced online update}
\begin{figure}
    \centering
     \setlength{\abovecaptionskip}{-0.2 cm}
\setlength{\belowcaptionskip}{-0.5 cm}
\includegraphics[height=0.2\textwidth]{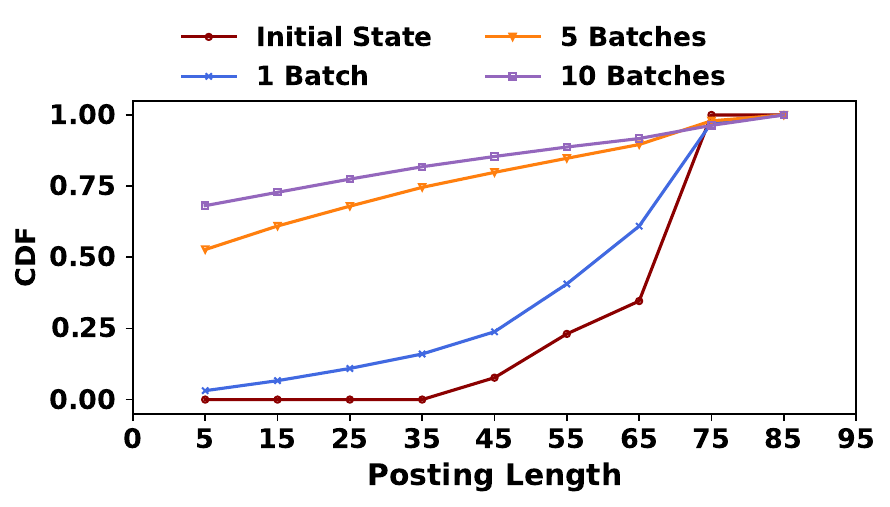}
    \caption{The posting distribution of different update batches in SPFresh. The merge threshold is set to 10, and the split threshold is set to 80. The test dataset is the Argoverse 2 motion forecasting dataset \cite{Argoverse2}.}  
    \label{fig:posting_cdf}
    
\end{figure}

We introduce non-trivial optimization of update operations upon SPFresh, and we incorporate additional branches to identify uneven cases and achieve a more balanced data distribution. Frequent updates can usually lead to a data shift, since new vectors will influence the centroids of the postings. Inexact centroids disturb search accuracy, and imbalanced distribution limits the search region, especially for the on-disk indices. SPFresh employs its new update policy (details in Section \ref{sec:update policy}) to figure out the centroid shift problem and support dynamic centroid update. However, it suffers from producing plenty of small postings during its newly proposed split processes in the streaming workloads. Figure \ref{fig:posting_cdf} shows the cumulative distribution function (CDF) of posting length after filtering the deleted postings. It is obvious that the ratio of small postings increases as more batches are fed. SPFresh faces this issue because the restrictions for triggering its proposed operations are too strict:
\begin{itemize}
[leftmargin=1em]
    \item A merge job is triggered only when a posting is accessed during a certain search task and satisfies the condition that its length is less than the configured value.
    \item A split job is triggered only when a posting is accessed during a certain insert task and satisfies the condition that its length is greater than the configured value. 
\end{itemize}

\begin{algorithm}[t]
\small
    \caption{BalanceSplit($I, P_i, l_{max},f$)}
    \label{algo:balance split algorithm}

    \SetKwInOut{Input}{Input}\SetKwInOut{Output}{Output}
    \Input{The index $I$, the original posting $P_i$, the configured split threshold $l_{max}$ and balance factor $f$.}
    \Output{two sub-postings $P_{i1}$ and $P_{i2}$.}
    \BlankLine
    
    $P_i' \leftarrow $ Filter deleted vectors in $P_i$\;
    \uIf{$|P_i'| < l_{max}$}
    {
    Replace $P_i$ with $P_i'$ on the disk\;
    $P_{i1} \leftarrow \emptyset; P_{i2} \leftarrow \emptyset$\;}
    \uElse{
        $P_{i1},P_{i2} \leftarrow $ Run $k$-Means algorithm on $P_i'$ where $k=2$\;

        \uIf{$min(|P_{i1}|,|P_{i2}|)< f * (|P_{i1}|+|P_{i2}|)$} {
            $P_{min} \leftarrow $The smaller posting between $P_{i1}$ and $P_{i2}$\;
            $P_{max} \leftarrow $The bigger posting between $P_{i1}$ and $P_{i2}$\;
            \ForEach{$p \in P_{min}$}{
                \uIf{$\exists P_j \in I, d(c_j,p) < d(c_{max},p)$}{
                    Append $p$ to posting $P_j$ and persist $P_j$\;
                }
                \lElse{Append $p$ to posting $P_{max}$}
            }
            $P_{i1} \leftarrow P_{max}; P_{i2} \leftarrow \emptyset $\;
        }
        Mark $P_i'$ as deleted\;
    }
    Persist the non-empty sets between $P_{i1}$ and $P_{i2}$\;
    \Return $P_{i1},P_{i2}$\;

\end{algorithm}

\par In such context, a split job, where the original posting is split into two smaller ones, may produce small postings and store them on disk since the merge jobs cannot be triggered. Based on this experiment and the analysis, we design a new module, called \textit{Balance Detector}, to monitor the posting distribution and reduce the ratio of small postings, such that the index is more balanced and achieves better search accuracy.

\par \textit{Balance Detector} consider achieving a more balanced distribution from the following two aspects:
\begin{itemize}
[leftmargin=1em]
    \item \textbf{Relaxing Restrictions.} It is necessary to incorporate additional branches for inspecting small postings. However, traversing each posting on the disk to check its length is also a resource-intensive operation. Considering this trade-off, in order to relax the restrictions, \textit{Balance Detector} records each posting length in memory and periodically examines the illegal postings in the background. Only postings that meet the conditions will have their complete data read from the disk and trigger the split or merge jobs. 
    \item  \textbf{Identifying Root.} The initial index stays in a relatively balanced state as Figure \ref{fig:posting_cdf} shows. The streaming update requests flood in and cause the centroid shift, which triggers lots of split jobs. However, some split jobs produce some small postings and merge jobs cannot access them, which is the root of the imbalanced distribution. \textit{Balance Detector} can identify these uneven cases where the small postings are produced in each split job. Algorithm \ref{algo:balance split algorithm} is an example for split jobs, which describes the details about how \textit{Balance Detector} influences the imbalanced split process. Lines 1-4 filter some deleted vectors from $P_i$ and check whether the filtered posting $P_i'$ still exceeds the threshold. The job will be abandoned if it fails to satisfy this condition. A $k$-Means clustering algorithm \cite{chen2021spann} is used for obtaining two smaller sub-postings on the vectors in the original posting. Lines 7-15 examine if the clustering process produces a small posting $P_{min}$ and search nearer existing posting of each vector $p \in P_{min}$. If so, vector $p$ will be added to its nearer posting $P_j$. Otherwise, $p$ will be appended to $P_{max}$, since they originally stay in the same cluster and share a high similarity. The original posting $P_i$ is then marked as deleted and new sub-postings that are not empty are persisted on disk. Merge jobs follow a similar idea.
    
\end{itemize}

\begin{table}[t]

\centering
\setlength{\abovecaptionskip}{-0.01 cm}
\caption{Static ANN datasets.}
\label{tab:Synthetic ANN datasets}

\begin{tabular}{cccc}
\hline
\textbf{Datasets} & \textbf{Dimension} & \textbf{Base Vectors} & \textbf{Query Vectors}  \\ \hline
SIFT1M  & 128 & 1,000,000 & 10,000  \\
Cohere1M  & 768 & 1,041,873 & 1,000  \\
GLOVE1M & 200 & 1,183,514 & 10,000    \\
\hline
\end{tabular}
\end{table}

\begin{table}[t]
\centering
\setlength{\abovecaptionskip}{-0.01 cm}
\caption{Thread allocation of each method.}
\label{tab:thread allocation in streaming update workload}
\begin{tabular}{|c|c|c|c|}
\hline
                  & FreshDiskANN & SPFresh & {\systemname} \\ \hline
Foreground Update & 4            & 1       & 1    \\ \hline
Background Update & 6            & 4       & 4    \\ \hline
Search            & 4            & 4       & 4    \\ \hline
\end{tabular}
\end{table}

\begin{figure*}[htbp]
\centering
 \setlength{\abovecaptionskip}{-0.01 cm}
\setlength{\belowcaptionskip}{-0.4 cm}

\includegraphics[width=\textwidth]
{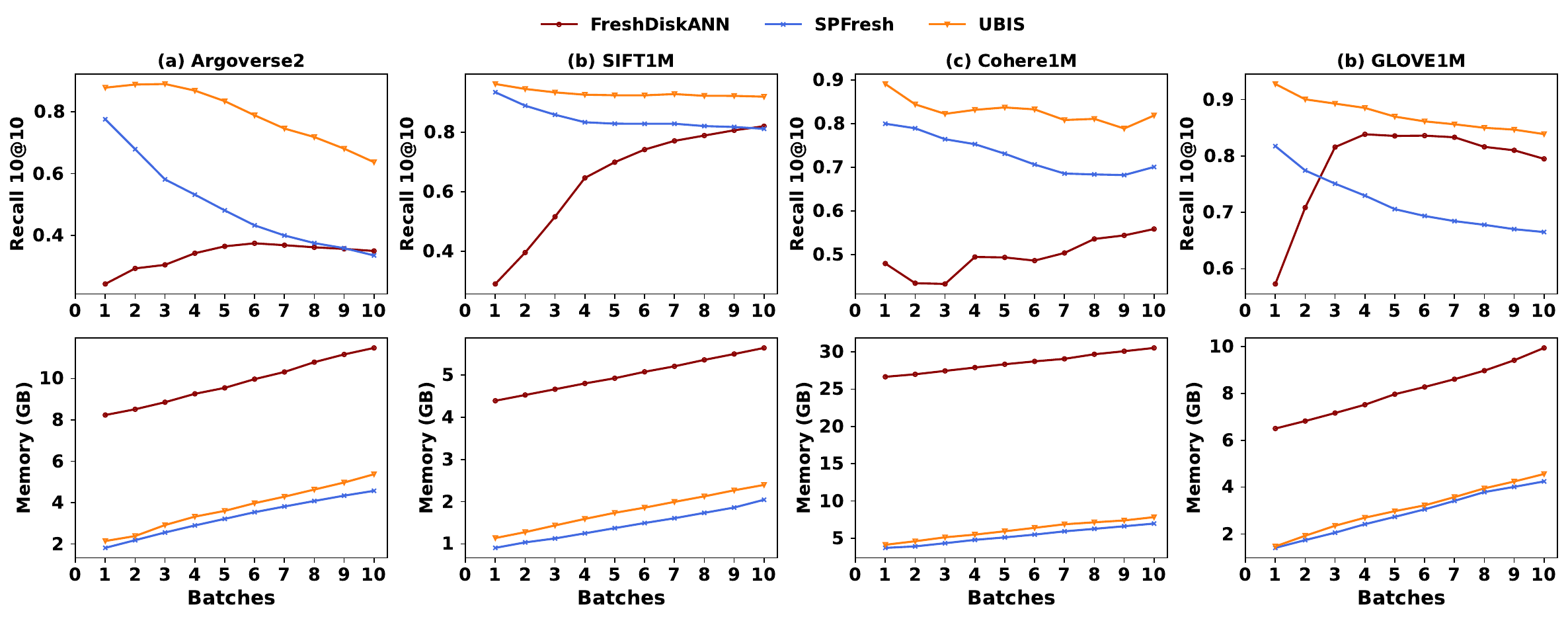}

\caption{The search accuracy and memory usage in \textit{streaming update} workload. {\systemname} achieves a higher recall rate compared with other methods.}
\label{fig:streaming update argoverse2 recall}
\end{figure*}

\section{Experiments}
\label{sec:experiments}

In this section, we will conduct experiments using two different common update workloads on both real-world datasets and synthetic datasets to demonstrate the efficiency of {\systemname}:
\begin{itemize}
[leftmargin=1em]
    \item We will update the initial index with update streams divided into multiple batches, and evaluate the index quality after each batch. Experimental results show that {\systemname} achieves about 77\% higher search accuracy and 45\% higher TPS on average (Section \ref{sec:streaming update experiment}).
    \item We will update the initial index with all fresh vectors at once to evaluate the overall performance. Experimental results demonstrate that {\systemname} performs about 16\% higher accuracy and about 52\% TPS on average, compared to existing indices (Section \ref{sec:full update experiment}).
    \item We also study the influence of both the fore-background thread number ratio and the newly-introduced balance factor in the same streaming update workload, to achieve better performance of {\systemname} (Section \ref{sec:parameter study}).
\end{itemize}

\subsection{Experiment Setups}\label{sec:experiment setup}

\myparagraph{Datasets}We utilize the autonomous driving dataset and other widely used ANN datasets to evaluate the update performance.
\begin{itemize}[leftmargin=1em]
    \item \textit{Dynamic Argoverse2 motion forecasting dataset (\textbf{Argoverse2})} \cite{Argoverse2}: This datasets contains about 1 million trajectories with real-world timestamps and each trajectory can be embedded to a 256-dimensional vector. Therefore, different trajectories can be sorted in chronological order naturally.
    
    \item  \textit{Static ANN Datasets} \cite{ann_datasets}: 
    We also utilize widely used ANN datasets, which are shown in Table \ref{tab:Synthetic ANN datasets}. These datasets lack additional temporal information, so we need to simulate the update sequence of each batch. These vectors are sorted based on the Gaussian distribution, so the vector number in each batch is close to the average value. 
    
\end{itemize}

\myparagraph{Comparisons}We select the following ANNS indices that support update to compare the update performance, where they are the representatives of graph-based indices and cluster-based indices, respectively. 

\begin{itemize}[leftmargin=1em]
    \item  \textit{FreshDiskANN} \cite{singh2021freshdiskann}: It is a state-of-the-art graph-based index that supports incremental update in the HNSW index \cite{malkov2018efficient}. It receives the new vectors in the local memory index and merges them into the disk index periodically.
    \item \textit{SPFresh} \cite{xu2023spfresh}: It is built on the state-of-the-art cluster-based index SPANN \cite{chen2021spann}. It employs a LIRE protocol to support in-place incremental update in the hierarchical index.

\end{itemize}

\myparagraph{Workloads}We utilize the following workloads.
\begin{itemize}[leftmargin=1em]
    \item \textit{Streaming update}: We divide the sorted query set into multiple batches using a fixed batch range, and each batch corresponds to a valid base vector range according to our recorded position data. Due to the uniform distribution, the number of vectors in each batch is close to the total number divided by the number of batches. We feed these batches continuously to form the streaming update workload. After each batch is completed, we evaluate the metrics.
    \item \textit{Full update}: This workload is the classic form of updating data inputs. It attempts to append all the vectors to the initial index. The search process will be triggered when all the processes have finished. We measure the metrics to evaluate the overall performance of the final updated index.
\end{itemize}

\myparagraph{Metrics}We utilize the following metrics in our experiments.
\begin{itemize}[leftmargin=1em]
    \item \textit{Search accuracy}:  We use recall to evaluate search accuracy. 
    
    \item \textit{Search efficiency}: We leverage query per second (QPS) and measure P99 tail latency to quantify the search efficiency.
    
    \item \textit{Update efficiency}:  We utilize the update throughput to evaluate the update efficiency. The update throughput, called transaction per second (TPS), could be measured by the average number of update jobs in each batch in one second.

    \item \textit{Memory usage}: We record snapshots of memory usage of different architectures, in order to reflect the resource consumption when update jobs are executed.
    
\end{itemize}

\begin{figure*}[htbp]
\centering
\setlength{\abovecaptionskip}{-0.01 cm}
\setlength{\belowcaptionskip}{-0.4 cm}
\includegraphics[width=\textwidth]{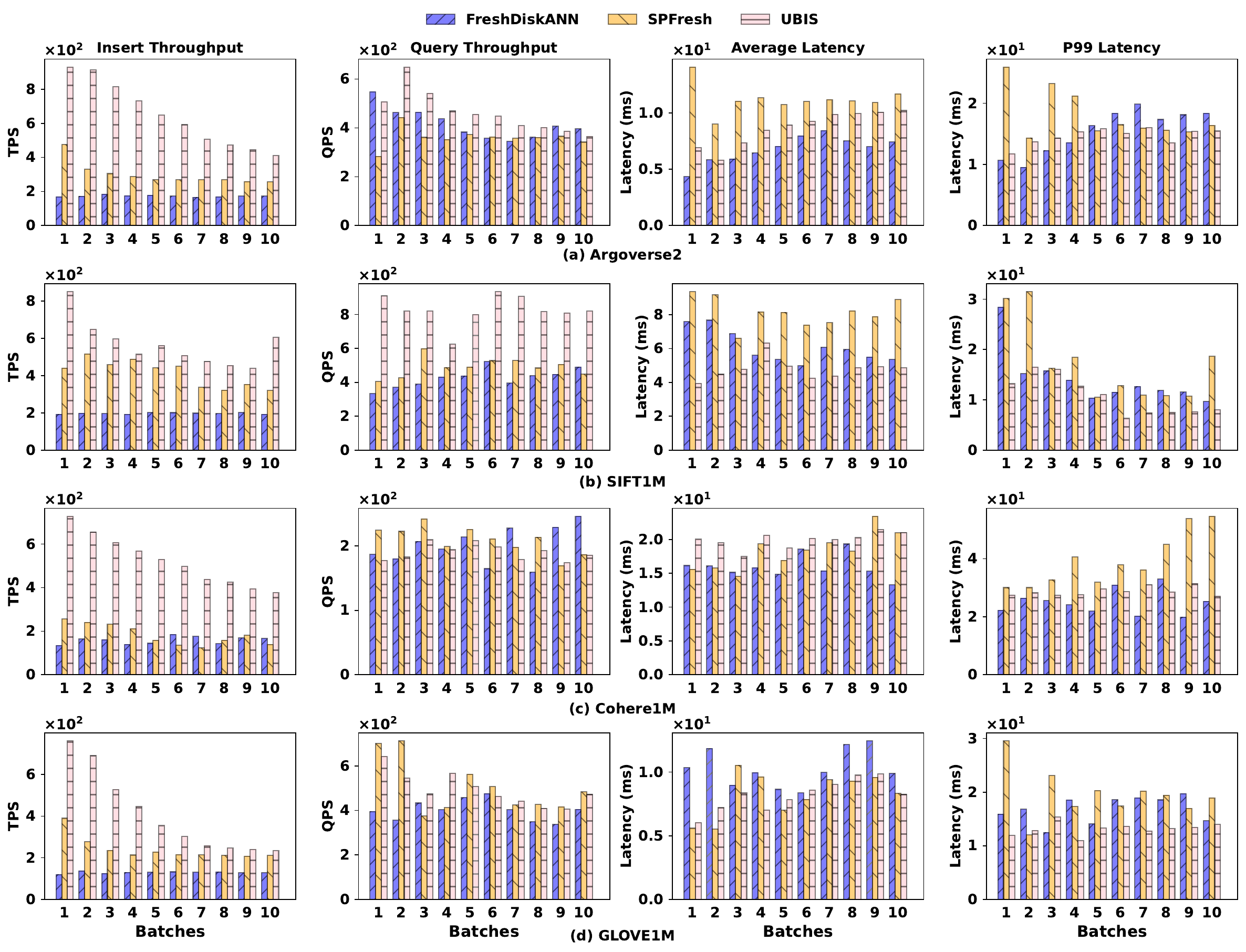}

\caption{The update and search efficiency in \textit{streaming update} workload. {\systemname} achieves the highest update throughput with little search performance decline.}
\label{fig:streaming update argoverse2 vector num}
\end{figure*}

\myparagraph{Configurations} We employ the same parameters on different workloads. Table \ref{tab:thread allocation in streaming update workload} shows the thread allocation for each method. As for FreshDiskANN, we increase the foreground thread number to improve its update efficiency. We set the maximum out-degree of the graph index in memory to 32 and that for index on disk to 64, which is the same as that in the official repository \cite{freshdiskann_repo}. For SPFresh and {\systemname}, we deploy the same parameters for splitting or merging postings, where we set the split threshold to 80, the merge threshold to 10. As for the key parameters for similarity search, in order to achieve close QPS and compare their recall in each batch, we set the size of the dynamic search candidate list to 40 for FreshDiskANN. The number of searches for the nearest postings is set to 32 for {\systemname} while it is set to 64 for SPFresh.

\myparagraph{Testbed}The experiments are conducted on a single server with an Intel Xeon Platinum 8352V processor and NVMe SSDs (Samsung SSD 990 PRO 2TB). We deploy all methods using a docker environment, which is limited to 16 vCPUs and 64 GB of memory on the Ubuntu 20.04 LTS operating system. The code is implemented in C++ and is open-sourced\footnote{\url{\repourl}}.

\begin{table*}[t]
\centering
\setlength{\abovecaptionskip}{0.1 cm}
\setlength{\belowcaptionskip}{-0.1 cm}
\caption{The overall performance in \textit{full update} workload. {\systemname} achieves the highest search accuracy and update throughput when facing large-scale fresh vectors, while it avoids sacrificing too much search efficiency.}
\label{tab:full update performance}

\resizebox{0.9\textwidth}{!}{%
\begin{tabular}{clclccccc}
\hline
\multicolumn{2}{c}{\multirow{2}{*}{\textbf{Datasets}}} & \multicolumn{2}{c}{\multirow{2}{*}{\textbf{Methods}}} & \multicolumn{5}{c}{\textbf{Metrics}}                                                        \\ \cline{5-9} 
\multicolumn{2}{c}{}                                   & \multicolumn{2}{c}{}                                  & Recall 10@10   & TPS              & Memory Usage (GB) & QPS              & P99 Latency (ms) \\ \hline
\multicolumn{2}{c}{\multirow{3}{*}{Argoverse2}}        & \multicolumn{2}{c}{FreshDiskANN}                      & 0.348          & 319.129          & 11.481            & \textbf{442.523} & \textbf{13.276}  \\
\multicolumn{2}{c}{}                                   & \multicolumn{2}{c}{SPFresh}                           & 0.583          & 325.520          & \textbf{5.872}    & 425.470          & 14.231           \\
\multicolumn{2}{c}{}                                   & \multicolumn{2}{c}{{\systemname}}        & \textbf{0.679} & \textbf{595.360} & 7.547             & 403.259          & 17.274           \\ \hline
\multicolumn{2}{c}{\multirow{3}{*}{SIFT1M}}            & \multicolumn{2}{c}{FreshDiskANN}                      & 0.469          & 456.966          & 5.726             & 388.575          & \textbf{10.968}  \\
\multicolumn{2}{c}{}                                   & \multicolumn{2}{c}{SPFresh}                           & 0.812          & 517.541          & \textbf{2.174}    & 458.540          & 19.916           \\
\multicolumn{2}{c}{}                                   & \multicolumn{2}{c}{{\systemname}}        & \textbf{0.906} & \textbf{627.469} & 3.227             & \textbf{464.196} & 14.838           \\ \hline
\multicolumn{2}{c}{\multirow{3}{*}{Cohere1M}}          & \multicolumn{2}{c}{FreshDiskANN}                      & 0.523          & 181.073          & 30.216            & \textbf{211.540}  & \textbf{27.713} \\
\multicolumn{2}{c}{}                                   & \multicolumn{2}{c}{SPFresh}                           & 0.708          & 169.610           & \textbf{6.934}    & 183.300          & 53.724           \\
\multicolumn{2}{c}{}                                   & \multicolumn{2}{c}{{\systemname}}        & \textbf{0.747} & \textbf{451.330}  & 7.695             & 181.036          & 33.931           \\ \hline
\multicolumn{2}{c}{\multirow{3}{*}{GLOVE1M}}           & \multicolumn{2}{c}{FreshDiskANN}                      & 0.352          & 218.059          & 9.817             & \textbf{352.372} & \textbf{19.849}  \\
\multicolumn{2}{c}{}                                   & \multicolumn{2}{c}{SPFresh}                           & 0.635          & 136.627          & \textbf{4.163}    & 319.201          & 25.136           \\
\multicolumn{2}{c}{}                                   & \multicolumn{2}{c}{{\systemname}}        & \textbf{0.818} & \textbf{360.852} & 5.461             & 343.090           & 22.239           \\ \hline
\end{tabular}
}

\end{table*}

\subsection{Streaming Update Experiment}\label{sec:streaming update experiment}

\myparagraph{Overall Experiment Results} Figure \ref{fig:streaming update argoverse2 recall} and Figure \ref{fig:streaming update argoverse2 vector num} present the overall evaluation of {\systemname} under \textit{streaming update} workload.
To summarize, {\systemname} achieves about 77\% higher search accuracy and 45\% higher TPS on average than the state-of-the-art SPFresh, which demonstrates the effectiveness of {\systemname}.

\myparagraph{Higher Search Accuracy}Figure \ref{fig:streaming update argoverse2 recall} shows that {\systemname} achieves higher search accuracy compared to other indices when processing \textit{streaming update} workload. We can see that {\systemname} reaches a more accurate state as more and more vectors of different batches are appended to the index structure, compared to other methods. 
\par FreshDiskANN fails to maintain a graph of great quality because streaming data destroys the navigability of the original index. As for SPFresh, \textit{streaming update} workload can lead to imbalanced postings on disk. SPFresh fails to handle the situation that the split process produces two extremely uneven postings, especially for high-dimensional data, due to its strict triggers. Therefore, some postings with extremely small sizes remain on disk and disturb the search process. {\systemname} can identify the imbalanced splitting case and maintain a more uniform posting distribution, which helps to prevent the postings' sizes from getting diverse.

\par FreshDiskANN is not suitable for situations where memory resources are limited. It needs much more memory to handle the same datasets compared to cluster-based index. {\systemname} utilizes a little additional memory for newly proposed \textit{Posting Recorder} and vector cache to temporarily store some fresh vectors, such that they can be quickly accessed.

\myparagraph{Better Update Efficiency}Figure \ref{fig:streaming update argoverse2 vector num} shows that {\systemname} achieves the highest update throughput in the evaluation, while our search efficiency is not sacrificed too much compared to the state-of-the-art methods, and even better in some cases.
\par {\systemname} behaves with a higher TPS than other methods in the same context. This is because {\systemname} does not require maintaining complex neighborhood relationships in FreshDiskANN, and eliminates lock design at the posting granularity in SPFresh. In addition, {\systemname} optimizes the expensive split and merge operations, which helps to accelerate these operations and reduce imbalanced cases. {\systemname} does not compromise search efficiency while it helps to improve the update throughput as Figure \ref{fig:streaming update argoverse2 vector num} shows.

\subsection{Full Update Experiment}\label{sec:full update experiment}

\myparagraph{Overall Experiment Results} Table \ref{tab:full update performance} shows the overall performance in the \textit{full update} workload. When fully updating the index structure, {\systemname} can reach about 16\% higher search accuracy and 52\% TPS on average than the other state-of-the-art indices, while it also maintains the same search efficiency.

\begin{figure}[t]
    \centering
    \setlength{\abovecaptionskip}{0 cm}
\setlength{\belowcaptionskip}{ 0cm}
    
    \includegraphics[width=\linewidth]{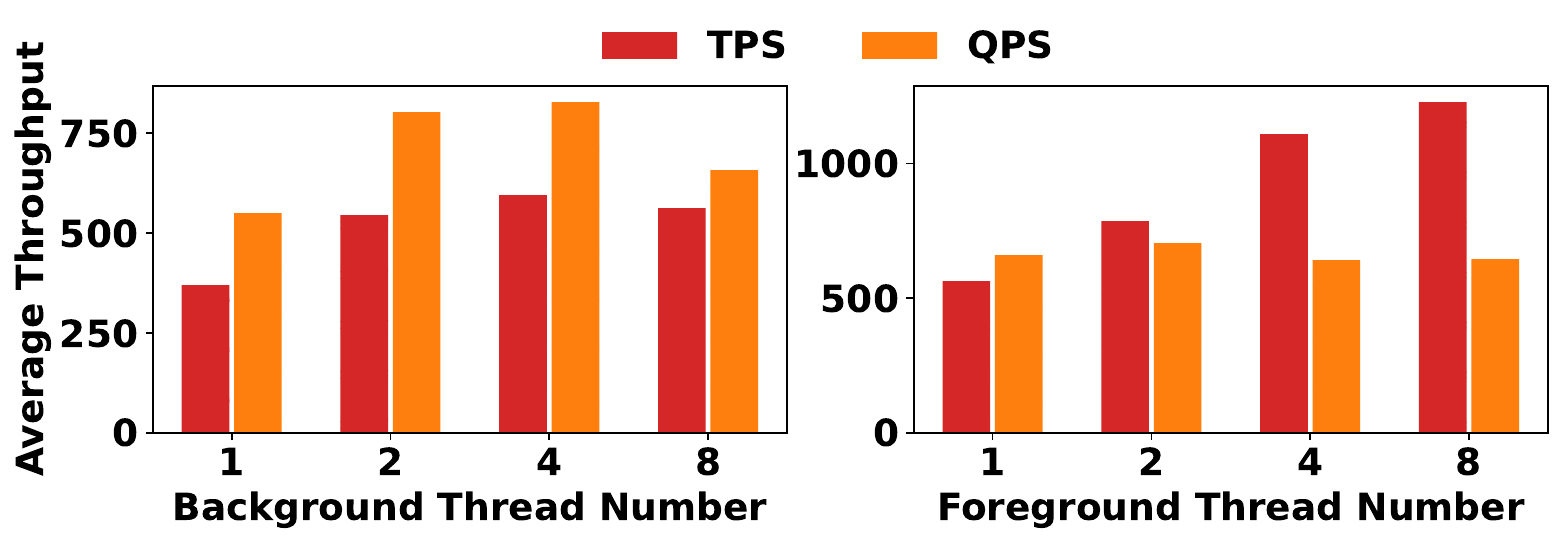}
    
    \caption{Foreground (background thread = 8) and background (foreground thread = 1) thread scalability in \textit{streaming update} workload on SIFT1M.}
    \label{fig:parameter study thread throughput}
\end{figure}

\myparagraph{More Reliable Search}As Table \ref{tab:full update performance} shows, {\systemname} outperforms the other indices in the aspect of search accuracy, which means {\systemname} can achieve a higher recall during the update process.

\par Although FreshDiskANN behaves the best QPS and P99 search latency in \textit{full update} workload, the accuracy is relatively low under the impact of large-scale fresh vectors. The fresh vectors lead to a lot of re-connections among neighbors, which destroys the navigability of the graph index. Therefore, FreshDiskANN fails to provide a reliable vector search. 

\par {\systemname} still performs the best search accuracy and update efficiency in \textit{full update} workload. {\systemname} leverages the update-friendly cluster-based index structure compared to FreshDiskANN, and it also optimizes the lock design and the expensive split operations compared to SPFresh. In addition, {\systemname} does not need to repeat the processes of searching the nearest posting, just like SPFresh. The \textit{Posting Recorder} can indicate the nearest candidates without traversing the index. The vector cache also provides fast access to some fresh vectors. To summarize, {\systemname} achieves about 16\% higher search accuracy and 52\% TPS on average compared to SPFresh, which shows the updatable ability of {\systemname}.

\myparagraph{Efficient Updatable Structure}Table \ref{tab:full update performance} shows that {\systemname} behaves the best update efficiency. 
It can receive more fresh vectors and obtain higher TPS in a more effective way, due to our proposed fine-grained lock and \textit{Posting Recorder}. These fresh vectors can be directly appended to them, since they originally stayed in the same group before split operations and are more likely to be the candidates. Table \ref{tab:full update performance} also shows that {\systemname} does not compromise search efficiency too much. 
{\systemname} achieves significantly higher accuracy and TPS than other methods, while maintaining a comparable QPS and latency. 

\par In summary, {\systemname} achieves better search accuracy and update throughput in \textit{full update} workload. It does not compromise search efficiency and achieves the best QPS on SIFT1M.

\subsection{Parameter Studies}\label{sec:parameter study}

\begin{figure}
    \centering
\setlength{\abovecaptionskip}{-0.2 cm}
\setlength{\belowcaptionskip}{0 cm}
    
    \includegraphics[width=0.82\linewidth]{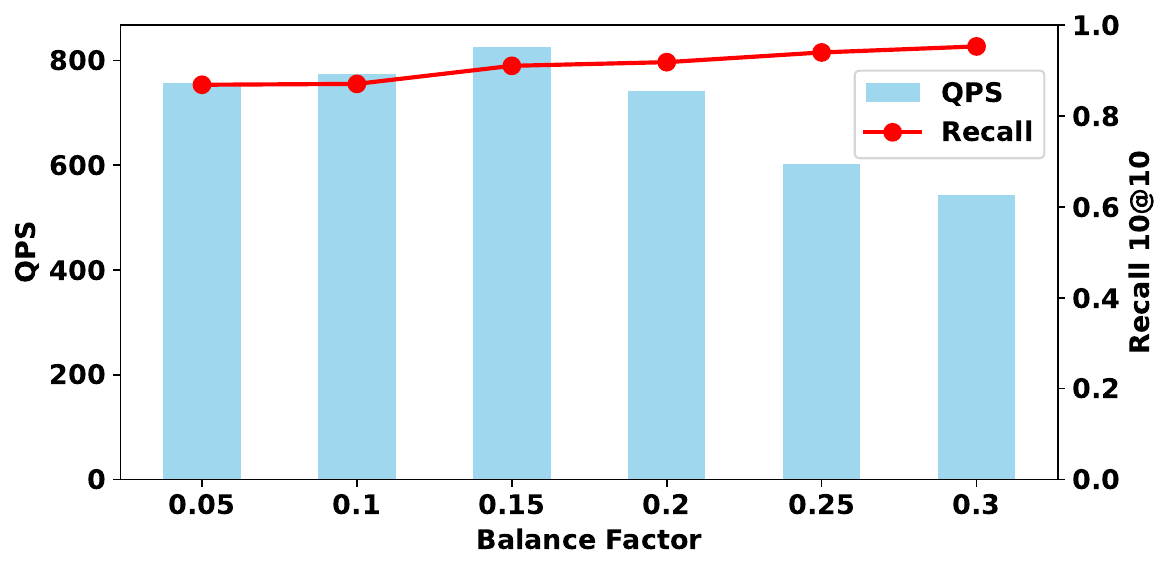}
    
    \caption{Balance detection threshold in \textit{streaming update} workload on SIFT1M. Although the recall increases as the balance factor increases, QPS decreases.}
    \label{fig:parameter study balance factor}
\end{figure}

\myparagraph{Overall Experiment Results} We aim to achieve better performance of {\systemname}. Figure \ref{fig:parameter study thread throughput} tells us that the ratio between foreground and background is best set at 1:4, to achieve higher TPS and higher QPS with less resource consumption. Figure \ref{fig:parameter study balance factor} shows that the balance factor should be set to 0.15, considering the trade-off between QPS and recall.

\myparagraph{Thread Number Ratio}This parameter represents the resource allocation and influences the throughput of {\systemname}, especially for TPS. In this experiment, we set the search thread number to a fixed 4. QPS is influenced due to our proposed vector cache design. Slow update processes may lead to the accumulation in the vector cache and result in lower QPS.
\par Figure \ref{fig:parameter study thread throughput} shows that {\systemname} achieves a better trade-off between TPS and QPS, when the ratio between the foreground and background is set to 1:4. Both TPS and QPS reach the peak when the foreground thread number is 1 and the background thread number is 4. The TPS and QPS decrease when setting the background thread number to 8 as the left part of Figure \ref{fig:parameter study thread throughput} shows. As for the right part, although TPS is improved when increasing the foreground thread number, QPS decreases. Therefore, it is better to set the foreground thread number to 2 when the background thread number is 8.

\myparagraph{Balance Factor}This parameter is used to identify the extreme splitting cases that may lead to imbalanced posting distribution. Therefore, the value of this parameter must be reasonable for assessing the imbalance. If the value is too low, some imbalanced cases are not recognized successfully. Conversely, reassign operations are triggered even for normal cases, unnecessarily increasing the number of vectors stored on disk and in the vector cache, which may degrade QPS.

\par Figure \ref{fig:parameter study balance factor} displays the results of different values of the balance factor in the \textit{streaming update} workload. As the balance factor increases, the recall also keeps the same rising trend. Although recall is slightly improved, QPS suffers from a decline because some vectors are reassigned to existing postings or to the vector cache in memory, which increases the search latency and reduces QPS. Based on this, we decide to select 0.15 as the default value of the balance factor.

\section{Conclusions}

In this paper, we focused on the complex cluster-based index streaming update problem, which aimed to upgrade the internal structure using fresh data while maintaining a stable search accuracy. This problem was meaningful for many real-world services that required a real-time search in evolving datasets. We proposed {\systemname} to support stable streaming search and update for cluster-based index at high frequency, such that the $k$-NN search could achieve higher search recall and TPS without compromising search efficiency. Experiments on real-world datasets demonstrated that {\systemname} was more effective than state-of-the-art indices, achieving a higher recall and TPS in streaming update workloads. Future work can focus on enhancing the update ability in distributed architectures.

\section*{Acknowledgment}
This work was supported by the National Key R\&D Program of China (2023YFB4503600), National Natural Science Foundation of China (62202338),  and the Key R\&D Program of Hubei Province (2023BAB081).

\newpage
\balance
\bibliographystyle{ieeetr}

\bibliography{references}

\end{document}